\documentclass[3p]{elsarticle}

\usepackage{lineno,hyperref}
\makeatletter
\def\ps@pprintTitle{%
   \let\@oddhead\@empty
   \let\@evenhead\@empty
   \def\@oddfoot{\reset@font\hfil\thepage\hfil}
   \let\@evenfoot\@oddfoot
}
\makeatother

 \usepackage{booktabs}
 \usepackage{ltablex,booktabs}
 \usepackage{lineno}
\usepackage{caption}
\usepackage{graphicx}
\usepackage{graphics}
\usepackage{latexsym}
\usepackage{amssymb}
\usepackage{verbatim}
\usepackage{amsmath}
\usepackage{amsthm}
\usepackage{amsfonts}

\usepackage{xcolor,colortbl}

\definecolor{Gray}{gray}{0.85}
\definecolor{LightCyan}{rgb}{0.88,1,1}
\usepackage{array, boldline, makecell, booktabs}

\usepackage{xcolor}
\usepackage{mdframed}
\usepackage{titletoc}
\usepackage{etoolbox}
\usepackage{array,multirow}
\usepackage{soul}
\newtheorem{Theorem}{Theorem}%[section]

 \biboptions{comma,square}
\biboptions{}
\usepackage{booktabs} 
\usepackage{colortbl} 
\usepackage{xcolor} 
\usepackage{xfrac}
\usepackage{caption}
\usepackage{xcolor}
\usepackage{listings}

\usepackage{lipsum}

\usepackage{mathpazo}
\usepackage{caption}
\usepackage[english]{babel}
\usepackage[utf8]{inputenc}
\usepackage{tabulary}
\usepackage{colortbl}
\usepackage[utf8]{inputenc}
\usepackage[english]{babel}
 
\usepackage{multicol}
\usepackage{colortbl}

\usepackage{mathpazo}
\usepackage{booktabs,caption,array,ragged2e}
\newcolumntype{C}[1]{>{\centering\arraybackslash}p{#1}}
\newcolumntype{L}[1]{>{\RaggedRight\arraybackslash}p{#1}}

\usepackage[utf8]{inputenc}
\usepackage[english]{babel}
\usepackage{amsthm}
\newtheorem*{remark}{Remark}

\usepackage{mathpazo} 
\usepackage{booktabs,array,ragged2e,dcolumn}
\usepackage{array}
\makeatletter
\newcommand{\thickhline}{%
    \noalign {\ifnum 0=`}\fi \hrule height 1pt
    \futurelet \reserved@a \@xhline
}
\newcolumntype{"}{@{\hskip\tabcolsep\vrule width 1pt\hskip\tabcolsep}}
\makeatother
\usepackage{color, colortbl}

\usepackage[first=0,last=9]{lcg}

\usepackage[T1]{fontenc}
    \usepackage{makecell,tabularx}

\hypersetup{hidelinks=true}

\begin{document}

\begin{frontmatter}

\title{On the security  of the  hierarchical attribute based encryption scheme proposed by Wang \textit{et al.} }



\author{Mohammad Ali}
\ead{mali71@aut.ac.ir}
\author{Javad Mohajeri} %\corref{mycorrespondingauthor}}
\ead{mohajer@sharif.ir}
\author{Mohammad-Reza Sadeghi}
\ead{msadeghi@aut.ac.ir}

%\cortext[mycorrespondingauthor] %{Corresponding author}
%\ead{mohajer@sharif.ir}

\address[mymainaddress]{ Department of Mathematics and Computer Science, Amirkabir University of Technology, Tehran, Iran. }
\address[mysecondaryaddress]{Electronic Research Institute, Sharif University of Technology, Tehran, Iran }
\begin{abstract}
 Ciphertext-policy hierarchical attribute-based encryption (CP-HABE) is a promising cryptographic primitive for  enforcing the  fine-grained access control with scalable key delegation and user revocation mechanisms on the outsourced encrypted data in a cloud. Wang \textit{et al.} (2011)   proposed the first  CP-HABE scheme  and showed that the scheme is  semantically secure in the random oracle model \cite{44,5}.   Due to  some weakness in its key delegation mechanism, by presenting two  attacks, we demonstrate  the  scheme does not offer any confidentiality and fine-grained access control. In this way, anyone who has just one attribute can  recover any outsourced encrypted data in the cloud. 
\end{abstract}

\begin{keyword}
\texttt{Cloud computing, Hierarchical attribute-based encryption, Fine-grained access control}
\end{keyword}

\end{frontmatter}
\section{Introduction}
Attribute-based encryption (ABE) scheme \cite{1}  is a one-to-many cryptographic primitive which provides confidentiality and fine-grained access control over the outsourced encrypted data, simultaneously. It provides access control on the shared data  by specifying an access structure over the ciphertexts or data users' secret-keys. According to the position of the access structure, this  cryptographic primitive  can be divided into two categories; key-Policy ABE (KP-ABE) \cite{2} and ciphertext-policy ABE (CP-ABE) \cite{3}. In a KP-ABE the access structure is embedded in data users'  secret-keys   by the key-generator authority and each ciphertext is labeled by a set of descriptive attributes. A data user can decrypt the ciphertext if and only if  the user's access structure is satisfied by the ciphertext's attributes.   While, in a CP-ABE the access structure is embedded in the ciphertext by the data owner, and considering attributes of each data user, his/her secret-keys are  issued by the key-generator authority. A data users can recover an encrypted data if and only if his/her  attributes satisfy the access structure of the ciphertext. 

In an ABE scheme data users have to make queries to the key generator authority for their secret-keys. However, this can make  some problems in the scalability and flexibility of the system when a large number of data users want to get their secret-keys, simultaneously.

In order to address the scalability problem,  Wang \textit{et al.} proposed a CP-hierarchical ABE (CP-HABE) scheme \cite{5}, by combining a hierarchical identity based encryption (HIBE) scheme \cite{4} and a CP-ABE \cite{3} scheme. By partitioning the universal attribute set to some disjoint subsets, they considered several key generators   that each of them administers one of the subsets.  In this scheme  for each attribute, the data user just can  get the corresponding   secret-key from the key generator that manage a subset  which contains the mentioned  attribute. 
 
After that, this idea had been used  in several ABE schemes.  Wan \textit{et al.} proposed a Hierarchal attribute set-based  encryption (HASBE) \cite{22}, by combining the notion of HABE and the CP-ASBE scheme proposed by Bobba \textit{et al.} \cite{55}. Li \textit{et al.} \cite{7} proposed a multi-authority access control system with efficient  key delegation and user revocation mechanisms.  Using outsourcing technique, they significantly decrease the computational cost in the user side. Liu \textit{et al.} \cite{6} proposed a time-based proxy re-encryption scheme, by combining an HABE scheme and a proxy re-encryption scheme \cite{66,666}, with a wide flexibility in user revocation mechanism. In this scheme, data owner can be off-line along the user revocation phase. Huang \textit{et al.} \cite{8} proposed a data collaboration scheme, by using HABE model in the key delegation mechanism. As \cite{7}, data outsourcing has been used to reduce the data user's computational cost. 

 Although,  it has been proved that the CP-HABE scheme proposed by Wang \textit{et al.} \cite{5} is  semantically secure  in the random oracle model, we showed that this scheme is fully insecure according to the given security definition  in \cite{5}. The scheme has some obvious drawbacks in its key delegation mechanism which enables a malicious  user to decrypt all the shared encrypted  data in the cloud with just one attribute.

The rest of this letter is organized as follows:   Some necessary basic concepts will be reviewed in Section \ref{preliminaries}.    We introduce CP-HABE scheme proposed by  Wang \textit{et al.} \cite{5} and  its security definition  in Section \ref{on the security}. In Section \ref{Attack}  we give two  attacks on the scheme that  both of them break  the security of the scheme  with probability $1$.  The conclusion of the paper is presented in Section \ref{conclusion}.

\section{preliminaries} \label{preliminaries}
In this section, we introduce some required definitions and hardness assumptions.
\subsection{Bilinear map}
Consider two cyclic groups $G_1$ and $G_2$  of a prime order $p$. Suppose that  $P_0$ is a generator of $G_1$. A function $\hat{e}:{G_1} \times {G_1} \to {G_2}$ is a bilinear map if it has the following properties: 
\begin{enumerate}
\item
Non-degeneracy: $\hat{e}(P_0,P_0)\ne 1$.
\item
Bilinearity: $\hat{e}(aP_1,bP_2)=\hat{e}(bP_1,aP_2)=\hat{e}(P_1,P_2)^{ab}$, for any $a,b \in \mathbb{Z}_p$ and $P_1,P_2\in G_1$ .
\item
Computability: there is a polynomial time algorithm which compute $\hat{e}(P_1,P_2)$, for any $P_1, P_2 \in G_1$.
\end{enumerate}
Consider two  cyclic groups $G_1$, $G_2$ of prime order $p$, a bilinear map  $\hat{e}:{G_1} \times {G_1} \to {G_2}$, and a random generator $P_0 \in G_1$.  The \textbf{Bilinear Diffie–Hellman (BDH)} problem  is to compute $\hat{e}(P_0,P_0)^{abc}$ for three given elements $aP_0,bP_0,cP_0\in G_1$, where  $a,b$, and $c$ are three uniform elements of $\mathbb{Z}_p$.

 \subsection{Access structure} \label{Access structure} 
Consider a universal attribute set $\mathbb{U}=\left\{ {a_1},\ldots,{a_n}\right\}$. Each nonempty subset  $\mathbb{A}$ of $2^{\mathbb{U}}$ is called an  access structure  on $\mathbb{U}$.  For an access structure $\mathbb{A}$, any set in $ \mathbb{A}$ is called an authorized set of attributes and the other ones  are called unauthorized sets.

Any   access structure $\mathbb{A}$ can be specified by a logical proposition $\mathop  \vee \limits_{i = 1}^N C{C_i}$, where each $CC_i$, $i=1,\ldots,n$, is a conjunction clause of some attributes. For example, the access structure $\mathbb{A}=\left\{ {\left\{ {{a_1},{a_2}} \right\},\left\{ {{a_1},{a_3}} \right\},\left\{ {{a_2},{a_3},{a_4}} \right\}} \right\}$ corresponds to the logical proposition  $\mathop  \vee \limits_{i = 1}^3 C{C_i}=({a_1} \wedge {a_2}) \vee ({a_1}\wedge {a_3}) \vee ({a_2}\wedge{a_3}\wedge{a_4})$. For simplicity, $\mathbb{A}=\mathop  \vee \limits_{i = 1}^N C{C_i}$ is used for indicating an access structure. This type  of presentation is called disjunctive normal form (DNF).

\section{CP-HABE proposed by Wang  \textit{et al.} \cite{5}.} \label{3} 
 \label{on the security}
In this section, we first introduce  system model of the  Wang's scheme, then the applied algorithms of this system are introduced in detail. After that,  semantic security definition for a CP-HABE scheme which is  proposed by Wang et al. \cite{5} will be presented.
\subsection{Model definition  and constructions  } \label{wang}
In the CP-HABE scheme proposed by Wang \textit{et al.} \cite{4}   the disjunctive normal form (DNF) is used for expressing the access control policy   and  a hierarchical key generation and user revocation model is applied   to provide   scalable and flexible mechanisms. Moreover, in this scheme each domain authority manages  a number of disjoint attributes. 
 
This system consists of five  entities: the root master (RM),  the cloud service provider (CSP),  data owners, the domain authorities, and  data users. The RM is responsible for generating the global public parameters  and  master keys for domain authorities at the first level.  The  cloud service provider's role is to let a data owner to  store its data and share  them with some data users.  The role of data owner is  determining an access structure for his/her own data, encrypting  the data under it,  and outsourcing the encrypted data in the cloud. The domain authorities   generate attribute secret-keys for some of the entities (data users or domain authorities) which stay on the next level. Data users can decrypt the outsourced encrypted data  using their attribute  secret-keys.

 In this scheme,  the  applied  key generation algorithms, named  \textbf{CeateDM} and \textbf{CreateUser} adopt a hierarchical  approach. First, the RM  generates global public parameters of the system by the \textbf{Setup} algorithm  and then generates the master secret-key of the domain authority in the first level, using the \textbf{CeateDM} algorithm. After that, some domain authorities run the \textbf{CeateDM} algorithm and   generate   master secret-keys of the domain authorities in its children. Also, the domain authorities in the last level  generate identity secret-keys and attribute secret-keys of  the authorized  data users, using the \textbf{CreateUser} algorithm. When a  data owner wants to outsource some data to the cloud, he/she should define an access structure and encrypt his/her data under the access structure using  the \textbf{Encrypt} algorithm.  Each data user can access to an outsourced encrypted data by running the \textbf{Decrypt} algorithm if and only if his/her attributes satisfy the access structure corresponding to the encrypted data.  
 
 In this scheme, it is assumed that each domain authority $DM_i$, data user $u$, and attribute $a$ in the universal attribute set has a unique public-key $PK_i$, $PK_u$ and $PK_a$, respectively.  The scheme can be described by  the following five algorithms: 
\begin{enumerate}

\item
\textbf{Setup:} This algorithm is run by the RM. It takes the security parameter $n$ as input and   picks a  large prime number $q$,   two cyclic groups $G_1$ and $G_2$ of order $q$, a bilinear map  $\hat{e}: G_1 \times G_1 \to G_2$,   a uniform  element $mk_0 \in \mathbb{Z}_q$,  three random oracle $H_1: \{0,1\}^{*} \to G_1$, $H_2:G_2 \to \{0,1\}^{n}$ and $H_A:\{0,1\}^{*} \to \mathbb{Z}_q$, and a random generator $P_0 \in G_1$. The algorithm outputs the master secret-key $MK_0=mk_0$ and system public parameters   $params = \left( {q,G_1,G_2,e,P_0,Q_0,H_1,H_2,H_A} \right)$, where $Q_0=mk_0P_0$.
\item
\textbf{CreateDM:} This algorithm is run by the root master or a domain authority as the parent. The inputs of the algorithm are the system public parameters  $params$, master secret-key $MK_i$ of the parent and the public-key of the domain authority $DM_{i+1}$, $PK_{i+1}$.  The output of the algorithm is the  $DM_{i+1}$'s master secret-key $M{K_{i + 1}} = \left( {m{k_{i + 1}},{H_{m{k_{i + 1}}}},{SK_{i + 1}},Q - tupl{e_{i + 1}}} \right)$, where $mk_{i+1} \in \mathbb{Z}_q$ is the index of the random oracle $H_{mk_{i+1}}:\{0,1\}^{*}\to \mathbb{Z}_q $, $P_{i+1}=H_{1}(PK_{i+1})$, $SK_{i+1}=SK_i+mk_iP_{i+1}$, $Q - tuple_{0}=(Q_0)$, and  $Q - tuple_{i + 1}=(Q-tupe_i,Q_{i+1}=mk_{i+1}P_0)$, for $i \ge 0$.
\item
\textbf{CreateUser:} When a data user $u$ makes a query to the domain authority $DM_i$ for a secret-key corresponding to an attribute $a$,  $DM_i$ checks whether the user is authorized for $a$ or not. If so, it runs this algorithm to generate the identity secret-key $SK_{i,u}=(Q-tupe_{i-1},mk_imk_uP_0)$ and attribute secret-key $SK_{i,a,u}=SK_i+mk_imk_uP_a$, where $mk_u=H_A(PK_u)$ and $P_a=H_{mk_i}(PK_a)P_0$.
\item
\textbf{Encrypt:}  This algorithm is run by a data owner. It takes public parameter $params$, a message $M$, an access structure $\mathbb{A}=\mathop  \vee \limits_{i = 1}^N C{C_i}=\mathop  \vee \limits_{i = 1}^N \mathop  \wedge \limits_{j = 1}^{n_i} {a_{ij}}$, and a set  of the corresponding   public-key  of the  attributes, $\left\{ {P{K_{{a_{ij}}}}:1 \le i \le m,1 \le j \le n} \right\}$.
Suppose that  all of  the attributes in $CC_i$ are covered by a specified domain authority $DM_{i_{t_i}}$. For each $1 \le i \le N$, consider the unique path  $(I{D_1},\ldots,I{D_{i{t_i}}})$ for $DM_1$ to the domain $DM_{it_i}$ .
 The algorithm  outputs  a ciphertext $CT = (\mathbb{A},V,{U_0},{U_{12}},\ldots,{U_{1{t_1}}},{U_1},\ldots,{U_{N2}},\ldots,{U_{N{t_N}}},\\{U_N})$,  where ${V = M \oplus H_2(\hat{e}({Q_0},r{n_A}{P_1}))}$ and $n_A$ is the  lowest common multiple (LCM) of $n_1,\ldots,n_N$, $r \in \mathbb{Z}_q$ is a uniform element, $U_0=rP_0$, ${U_i} = r\sum\limits_{j = 1}^{{n_i}} {P{a_{ij}}}$, and ${U_{ik}} = r{P_{ik}}$, for $i=1,\ldots,N$ and $k=1,\ldots,t_i$.
\item
\textbf{Decrypt:} A data user whose attributes satisfy the access structure $\mathbb{A}=\mathop  \vee \limits_{i = 1}^N C{C_i}=\mathop  \vee \limits_{i = 1}^N \mathop  \wedge \limits_{j = 1}^{n_i} {a_{ij}}$ of a ciphertext $CT$,  can run this algorithm and recover the corresponding message. Suppose that for an $i \in \{1,\ldots,n_i\}$, a data user has all the determined attributes in $CC_i$, then  the corresponding  message can be obtained  as follows:
\[V \oplus {H_2}(\frac{{\hat{e}({U_0},\frac{{{n_A}}}{{{n_i}}}\sum\limits_{j = 1}^{{n_i}} {S{K_{{i_{{t_i}}},u,{a_{ij}}}}} )}}{{\hat{e}(m{k_u}m{k_{i{t_i}}}{P_0},\frac{{{n_A}}}{{{n_i}}}{U_i})\prod\nolimits_{j = 2}^{{t_i}} {\hat{e}({U_{ij}},{n_A}{Q_{i(j - 1)}})} }}).\]
We refer the reader to \cite{5} for  more detail about this scheme. 
\end{enumerate}

\subsection{Security definition:} \label{Security definition}
Consider the following game: 

\begin{enumerate}
\item
\textbf{Setup}:
The challenger runs  \textbf{Setup} algorithm  and gives the system public parameters  to the adversary. 
\item
\textbf{Phase 1}:
First of all, challenger runs \textbf{CreateDM} algorithm, then the adversary $\mathcal A$   makes an arbitrary number of queries  for users' attribute secret-keys. For each data user $u$, once the adversary makes a query for the user's  secret-key corresponding to an attribute  $a$,  the challenger runs   \textbf{CreateUser} algorithm, and gives the requested secret-key to the adversary $\mathcal{A}$.
\item
\textbf{Challenge}:
When the adversary decides to terminate Phase 1, he/she gives two equal length messages $m_0$, $m_1$ and an access structure $\mathbb{A} $ to the challenger, where the set of specified attributes for any data users in Phase 1, dose not satisfied the access structure $\mathbb{A}$. Then, the challenger randomly chooses $b \in \left\{ {0,1} \right\}$, encrypts $m_b$  under the access structure, and  returns the encrypted message  to the adversary $\mathcal{A}$. 
\item
\textbf{Phase 2}:
The adversary is allowed to make more attribute secret-keys query, with the same  constraints   in the previous phases.
\item
\textbf{Guess}:
The adversary outputs a bit $b'\in \left\{ {0,1} \right\}$. It wins this game if $b=b'$.
\end{enumerate}
Let notation  $Succeed(\mathcal A)$ denotes the event that the adversary $ \mathcal A $ succeeds in the above game.
A CP-HABE scheme is semantically secure if  $\left| {P(Succeed(\mathcal A) - \frac{1}{2})} \right|$ is  a negligible function in term  of the security parameter, for each polynomial time adversary $\mathcal A$. 

In  Appendix A of \cite{5}, the semantic security of the CP-HABE scheme has been proved  based on the  hardness assumption of BDH problem, in the random oracle model . In the next section we will show that this scheme is  vulnerable against our two proposed  attacks.

\section{Security analysis of the CP-HABE scheme proposed by Wang \textit{et al.}} \label{Attack}
We  show that there are   two drawbacks in the key delegation mechanism of the CP-HABE proposed by Wang \textit{et al.} \cite{44,5}. Considering these drawbacks, a malicious data user with just one or two attributes can decrypt any outsourced encrypted data in the cloud. 

In the following, we  propose two non-adaptive  attacks  on the CP-HABE scheme. Each of them breaks  the semantic security of the scheme with probability $1$.

\begin{remark} \label{kkk}
For  an arbitrary domain $DM_{i{t_i}}$, let $(ID_1,\ldots, ID_{i{t_i}})$ be the unique path from $DM_1$ to $DM_{i{t_i}}$. Then, we have:
\begin{align} \nonumber
S{K_{i{t_i}}} &= S{K_{i({t_i} - 1)}} + m{k_{i({t_i} - 1)}}{P_{i{t_i}}}\\ \label{kk}
 &= S{K_1} + \sum\limits_{j = 1}^{{t_i} - 1} {m{k_{ij}}{P_{i(j + 1)}}}.
\end{align}
 
\end{remark}

\begin{Theorem} \label{lem}
For an arbitrary domain  $DM_{i{t_i}}$ with  $(ID_1,\ldots, ID_{i{t_i}})$, any user $u$ who has received his/her  identity secret-key $S{K_{i{t_i},u}}$ and obtained the secret-key of $DM_{i{t_i}}$,  $SK_{i{t_i}}$, can recover any outsourced encrypted data to the could.

\begin{proof}
Sine the user has received $S{K_{i{t_i},u}} = \left( {Q - tupl{e_{i{t_i} - 1}},m{k_{i{t_i}}}m{k_u}{P_0}} \right) = \left( {{Q_0},\ldots,{Q_{i({t_i} - 1)}},m{k_u}{Q_{i{t_i}}}} \right)$ and can obtain $mk_u=H_A(PK_u)$, he/she can calculate $Q_{i{t_i}}$ by multiplying $mk_u^{-1}=(H_A(PK_u))^{-1} \in \mathbb{Z}_q$ to the last component of $S{K_{i{t_i},u}}$, $m{k_u}{Q_{i{t_i}}}$. So, the user know $Q_{ij}$ for each $1 \le j \le t_i$. Let $CT = (\mathbb{A},V,{U_0},{U_{12}},\ldots,{U_{1{t_1}}},\\{U_1},\ldots,{U_{N2}},\ldots,{U_{N{t_N}}},{U_N})$ be   an arbitrary outsourced encrypted data. Then, since $V = M \oplus {H_2}(\hat{e}({Q_0},{n_A}r{P_1}))$ we have:

\begin{align} \nonumber
M &= V \oplus {H_2}(\hat{e}({Q_0},{n_A}r{P_1}))\\ \nonumber
& = V \oplus {H_2}(\hat{e}(m{k_0}{P_0},{n_A}r{P_1}))\\ \nonumber
& = V \oplus {H_2}(\hat{e}({n_A}r{P_0},m{k_0}{P_1}))\\ \nonumber
&= V \oplus {H_2}(\hat{e}({n_A}{U_0},S{K_1}))\\  
& = V \oplus {H_2}(\hat{e}({n_A}{U_0},S{K_1} + \sum\limits_{j = 1}^{{t_i} - 1} {m{k_{ij}}{P_{i(j + 1)}}}  - \sum\limits_{j = 1}^{{t_i} - 1} {m{k_{ij}}{P_{i(j + 1)}}} ))
\end{align}

From Equation \ref{kk}, we conclude that:
\begin{align} \nonumber
M &= V \oplus {H_2}(\hat{e}({n_A}{U_0},S{K_{i{t_i}}} - \sum\limits_{j = 1}^{{t_i} - 1} {m{k_{ij}}{P_{i(j + 1)}}} ))\\ \nonumber
& = V \oplus {H_2}(\hat{e}({n_A}{U_0},S{K_{i{t_i}}}).\hat{e}( - {n_A}{U_0},\sum\limits_{j = 1}^{{t_i} - 1} {m{k_{ij}}{P_{i(j + 1)}}} ))\\ \nonumber
& = V \oplus {H_2}(\hat{e}({n_A}{U_0},S{K_{i{t_i}}}).\prod\limits_{j = 1}^{{t_i} - 1} {\hat{e}( - {n_A}r{P_0},m{k_{ij}}{P_{i(j + 1)}})} )\\ \nonumber
& = V \oplus {H_2}(\hat{e}({n_A}{U_0},S{K_{i{t_i}}}).\prod\limits_{j = 1}^{{t_i} - 1} {\hat{e}( - {n_A}m{k_{ij}}{P_0},r{P_{i(j + 1)}})} )\\ \label{lll}
& = V \oplus {H_2}(\hat{e}({n_A}{U_0},S{K_{i{t_i}}}).\prod\limits_{j = 1}^{{t_i} - 1} {\hat{e}( - {n_A}{Q_{ij}},{U_{i(j + 1)}})} ),
\end{align}
  According to the assumption that the data user $u$ has obtained   $SK_{it_i}$, since he/she know  $Q_{ij}$, for $j=1, \ldots,t_i$, and the other involved parameter  in (\ref{lll}), the ciphertext can be decrypted by the user. 
\end{proof}
\end{Theorem}
%Here we present two attacks on the CP-HABE scheme proposed by Wang \textit{et al.} \cite{5} and we show that  according to the mentioned security definition  in the last section, the scheme is not secure. We first describe an attack which reveals an obvious drawback of the scheme in its key delegation mechanism. Then, we demonstrate that,  using this drawback  anyone  with just an attribute secret key  can decrypt any  encrypted data. Therefore, he/she can easily win the introduced  game   Section \ref{3} 
\subsection{Attack 1} \label{at1}

This attack shows that any user who has just one attribute administrated by a domain $DM_{it_i}$  can obtain $SK_{it_i}$. Therefore, from Theorem \ref{lem}, we get the user can recover any outsourced encrypted data to the cloud.

According to the Security definition presented in the last section. Let   a challenger  has run the \textbf{Setup} and \textbf{CreateDM} algorithms and $\mathcal A$ be   a polynomial time adversary  which is taken the  system public parameters $params$ generated by \textbf{Setup} algorithm.
	\begin{itemize}
		\item 
	Then  $\mathcal A$  picks just  an arbitrary attribute $a_0$ and authorized data user $u_0$ to the attribute and makes a query for the corresponding secret-key. The challenger runs the \textbf{CreateUser} algorithm and gives the requested secret-keys, $SK_{it_i,u_0}$ and  $SK_{it_i,a_0,u_0}=SK_{it_i}+mk_{it_i}mk_{u_0}P_{a_0}$,  to the adversary $ \mathcal{A} $. 
		\item 
		At  \textbf{Challenge} step, $\mathcal A$   gives two random equal length plaintexts $m_0$ and $m_1$, and a DNF access structure $\mathbb{A}= \mathop  \vee \limits_{i = 1}^N C{C_i}$ to the challenger, where $CC_1$   includes $a_0$ and also $|CC_1|>1$, therefore  $a_0$ does not satisfy $CC_1$. The challenger chooses a uniform bit $b \in \left\{ {0,1} \right\}$ and encrypts $m_b$ under an access strutter $\mathbb{A} = \mathop  \vee \limits_{i = 1}^N C{C_i}$. The generated ciphertext $CT$ is given to $\mathcal A$.  
		%\item 
	%\textcolor{blue}{In \textbf{Phase 2},   $\mathcal A$ does not make any secret-key query.} %chooses an attribute $a \in CC_k$ and   an arbitrary user $u$ who is authorized to $a$, $\mathcal A$ makes a    query   for the $u$'s  secret key  corresponding to  $a$. Challenger runs \textbf{CreateDM} and \textbf{CreateUser} algorithms and gives the identity secret key $SK_{i,u}=(Q-tupe_{i-1},mk_imk_uP_0)$ and attribute secret key $SK_{i,a,u}=SK_i+mk_imk_uP_a$ to the adversary.
	\item
	With no need to run Phase 2, in  Guess step, first, $\mathcal A$  calculates $H_{mk_{it_i}}(PK_a)$, then using the last component of $SK_{it_i,u_0}$,  $mk_{it_i}mk_{u_0}P_0$, it calculates $A_{a_0,u_0}=H_{mk_{it_i}}(PK_{a_0}).mk_{it_i}mk_uP_0=mk_{it_i}mk_{u_0}P_{a_0}$. Therefore, $\mathcal A$ can obtain $SK_{it_i}=SK_{it_i,a_0,u_0}-A_{a_0,u_0}$.
	
Now, from Theorem \ref{lem}, since the adversary has $SK_{u_0,i}$ and $SK_i$, he/she can decrypt $CT$ and get $m_0$. So the adversary can win the game with probability $1$

	\begin{Theorem}
	By using the described techniques in Attack 1, a polynomial time adversary $\mathcal{A}$ in a  non-adaptive manner  can  break the semantic security  of  the CP-HABE scheme proposed by Wang \textit{et al.} \cite{5} with probability $1$.
	\end{Theorem}

	\end{itemize}
	  Note that  using  Attack 1, an adversary with just one attribute secret-key, can decrypt any given ciphertext. 
	
		\subsection{Attack 2} \label{at2}
%Here we describe another attack which demonstrate the other drawback of the scheme proposed by Wang \textit{et al.} \cite{5}. 
	The attack shows that any user who has two attributes administrated by same domain can obtain master secret key of the domain and therefore from Theorem \ref{lem} he/she can decrypt any outsourced ciphertext to the cloud.  
	
As before, considering the security definition presented in Section 3, Let the 
  \textbf{Setup} and  \textbf{CreateDM} algorithms have been run by a challenger and system pubic-parameters are given to the adversary $\mathcal A'$.
\begin{itemize}
\item
The adversary picks two attributes $a_1$ and $a_2$, and a data user $u_0$. Then it makes two  queries for the corresponding  secret-keys.  The challenger runs \textbf{CreateUser} algorithm and gives the secret-keys $SK_{i,{u_0}}=(Q-tupe_{i-1},mk_imk_{u_0}P_0)$ ,  $SK_{i,a_1,{u_0}}=SK_i+mk_imk_{u_0}P_{a_1}$ and $SK_{i,a_2,{u_0}}=SK_i+mk_imk_{u_0}P_{a_2}$ to the adversary.
\item
  In \textbf{Challenge} step, the adversary $\mathcal A'$ gives  two equal length messages $m_0$ and $m_1$, and an access structure $\mathbb{A}= \mathop  \vee \limits_{i = 1}^N C{C_i}$ to the challenger, where $|CC_1|>2$.  The challenger uniformly chooses $b \in \{0,1\}$  and sends the generated ciphertext $CT$ corresponding to $m_b$ and $\mathbb{A}$ to the adversary. 
%  \item
%\textcolor{blue}{ The adversary does not make any query in \textbf{Phase 2}.}

\item
Without runing Phase 2, in \textbf{Guess} step, the adversary sets: 
\begin{align}\nonumber
{B_{a,{u_0}}} &= (SK_{i,a_1,{u_0}} - SK_{i,a_2,{u_0}}) \\ \nonumber
&= m{k_i}m{k_{u_0}}(P_{{a_1}} - P_{{a_2}}) \\ \nonumber
&= (H_{mk_i}(PK_{a_1})-H_{mk_i}(PK_{a_2}))  (mk_im{k_{u_0}}P_0).
\end{align}
 So,
 $C_{i,a_1,{u_0}}={H_{m{k_i}}}(P{K_{a_1}}){(H_{mk_i}(PK_{a_1})-H_{mk_i}(PK_{a_2}))^{ - 1}}{B_{a,{u_0}}}={H_{m{k_i}}}(P{K_{a_1}})mk_{u_0}m{k_i}P_0=m{k_{u_0}}m{k_i}{P_{a_1}}$
 can be obtained by the adversary.  Now, if $\mathcal {A'}$   sets  $S{K_i} = S{K_{i,{a_1},{u_0}}} - {C_{i,{a_1},{u_0}}}$, for Lemma \ref{lem} it can decrypt any outsourced ciphertext. Therefore the adversary can win the game with probability $1$. 

\begin{Theorem}
 Attack 2 enables a non-adaptive adversary to break the semantic security of the CP-HABE proposed by Wang \textit{et al.} with probability $1$.
\end{Theorem}
	\end{itemize}
	
\section{Conclusion} \label{conclusion}	

In this manuscript, we showed that the CP-HABE proposed by Wang \textit{et al.} \cite{5} is fully insecure. We provided two  attacks which break the scheme's security with probability $1$, that is contrary to the authors’ claim.  Moreover, it was  shown that any malicious  user who has  just one attribute can recover any outsourced encrypted data in the cloud.



\section*{References}
\begin{thebibliography}{1}

	\bibitem{1} Sahai, A. and Waters, B., 2005, May. Fuzzy identity-based encryption. In Annual International Conference on the Theory and Applications of Cryptographic Techniques (pp. 457-473). Springer, Berlin, Heidelberg.
	
		\bibitem{2}
		Goyal, V., Pandey, O., Sahai, A. and Waters, B., 2006, October. Attribute-based encryption for fine-grained access control of encrypted data. In Proceedings of the 13th ACM conference on Computer and communications security (pp. 89-98). Acm.
		
		\bibitem{3}
		Bethencourt, J., Sahai, A. and Waters, B., 2007, May. Ciphertext-policy attribute-based encryption. In Security and Privacy, 2007. SP'07. IEEE Symposium on (pp. 321-334). IEEE.
		
		\bibitem{44}  Wang, G., Liu, Q. and Wu, J., 2010, October. Hierarchical attribute-based encryption for fine-grained access control in cloud storage services. In Proceedings of the 17th ACM conference on Computer and communications security (pp. 735-737). ACM.
		
			\bibitem{5} Wang, G., Liu, Q., Wu, J. and Guo, M., 2011. Hierarchical attribute-based encryption and scalable user revocation for sharing data in cloud servers. computers \& security, 30(5), pp.320-331.
		
		\bibitem{4}
	C. Gentry and A. Silverberg. Hierarchical ID-Based
	Cryptography. In Proceedings of ASIACRYPT 2002,
	pages 548-566.
	
	
		\bibitem{22} Wan, Z., Liu, J.E. and Deng, R.H., 2012. HASBE: A hierarchical attribute-based solution for flexible and scalable access control in cloud computing. IEEE transactions on information forensics and security, 7(2), pp.743-754.


	
	\bibitem{55} Bobba, R., Khurana, H. and Prabhakaran, M., 2009, September. Attribute-sets: A practically motivated enhancement to attribute-based encryption. In European Symposium on Research in Computer Security (pp. 587-604). Springer, Berlin, Heidelberg.
	
	\bibitem{7} Li, Q., Ma, J., Li, R., Liu, X., Xiong, J. and Chen, D., 2016. Secure, efficient and revocable multi-authority access control system in cloud storage. Computers \& Security, 59, pp.45-59.
	
	\bibitem{6} Liu, Q., Wang, G. and Wu, J., 2014. Time-based proxy re-encryption scheme for secure data sharing in a cloud environment. Information sciences, 258, pp.355-370.
	
	\bibitem{66} M. Blaze, G. Bleumer, M. Strauss, Divertible protocols and atomic proxy cryptography, in: Proceedings of International Conference on the Theory and
	Application of Cryptographic Techniques (EUROCRYPT), 1998, pp. 127–144.
	

	\bibitem{666} M. Green, G. Ateniese, Identity-based proxy re-encryption, in: Proceedings of the International Conference on Applied Cryptography and Network
	Security (ACNS), 2007, pp. 288–306.

		
		
	\bibitem{8} Huang, Q., Yang, Y. and Shen, M., 2017. Secure and efficient data collaboration with hierarchical attribute-based encryption in cloud computing. Future Generation Computer Systems, 72, pp.239-249
			

	



 


	
	\end{thebibliography}
 \end{document}